# ENA imaging near Planetary Bodies: Interaction between Plasma, Exosphere and Surface

Yoshifumi Futaana
*Swedish Institute of Space Physics, Box 812, Kiruna SE 98128, Sweden*

## Abstract

Energetic Neutral Atom (ENA) imaging has been noticed as a powerful tool for remote sensing the plasma-neutral interaction in space. Particularly, the technique is used for investigation of space plasma near planetary bodies. Hear we provide a short review of recent low-energy ENA observations (up to ~1 keV) near Mars, Venus and the Moon.

## Introduction

It is frequently said that 99% of all known matters in the universe is in the plasma state (e.g. *Baumjohann and Treumann, 1997*). On the other hand, cis-planet environments are dominated by neutral matters such as their surfaces, dusts, atmospheres, and exospheres. When space plasma, which usually has higher energy than neutral matters, blows to the neutral matters, the plasma will be neutralized. Such neutralized matters with high energy are called energetic neutral atoms (ENAs).

Energetic Neutral Atoms (ENAs) are conventionally categorized by their energy. **Figure 1** shows typical energy ranges of low-, medium-, and high-energy ENAs. Note that typical energies of the neutral matters in space are much lower (in the order of 0.01 eV) compared with those of ENAs. The energy is conserved along the trajectory. Due to the neutrality, ENAs fly along a straight line, and therefore they are thus measured remotely. In the following review, we dedicate the low-energy ENA (LENA) near planetary bodies.

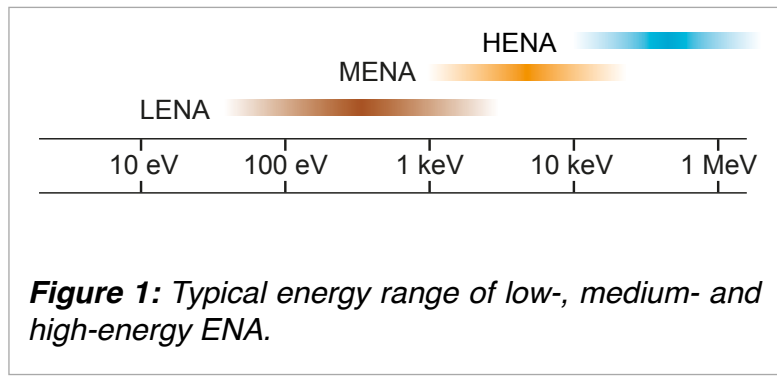

*Figure 1:* Typical energy range of low-, medium- and high-energy ENA.

## LENA generation

The LENA characteristics depend highly on the plasma and neutral environment near the object. The plasma environment is highly controlled by the existence of magnetic field and neutral atmosphere (*Barabash, 2012*). There are several LENA generation mechanisms, depending on the environment, to be addressed in terms of planetary ENA imaging.

The most famous ENA generation mechanism is the charge exchange process. When an ion received an electron by probability from a neighboring neutral atom, the primary ion becomes an ENA (and the neutral atom becomes an ion). During the charge exchange process, the energy loss of the primary ion is not very large (<10 eV for 1 keV ions); so that we can usually assume that the energy loss is negligible (e.g. *Basu et al.,1993*). Also, the directional change can be small enough to be ignored. This process occurs mainly when the target neutral matters are gas (atmosphere).

Interaction of ions with a surface is a big scientific field. However, in terms of ENA imaging in space for planets, two physical mechanisms are mainly addressed, namely sputtering and scattering. When an ion hits the surface, the energy and momentum is transferred to surface atoms, and a part of them can be released from the surface. This mechanism is sputtering. Ideal energy spectrum follows the famous Thompson-Sigmund spectrum (e.g. *Thompson, 1968; Betz and Wien, 1994; Futaana et al., 2006a; Wurz et al., 2007*). On the other hand, it is also possible that the primary ion hitting the surface is neutralized and back into space. This is the scattering process. For the scattering process, the energy loss is roughly several 10s %. For details, one may find a review by *Niehus et al. [1993]*. Indeed, both mechanisms may happen even for the interaction with atmosphere (e.g. *Luhmann et al., 1992; Futaana et al., 2006b*).

## ENA from Mars and Venus

Mars and Venus do not have intrinsic magnetic field. Therefore, the solar wind plasma can directly interaction with their atmospheres. Several ENAs have been theoretically predicted; solar wind and magnetosheath (shocked solar wind) ENAs by charge exchange, ionospheric origin ENAs by charge exchange, exospheric origin ENAs by sputtering, or scattered solar wind ENAs (see a review by *Futaana et al., 2011*).

Mars Express and Venus Express carried ENA detectors, called NPD (Neutral Particle Detector), as a part of ASPERA-3 and -4, respectively. Particular interests were addressed to the solar wind interaction with atmosphere. The energy transfer from the solar wind to atmosphere is one of the objectives of NPD. *Futaana et al., [2006b]* detected for the first time hydrogen ENA emission from the Martian upper atmosphere. We interpreted the emission as backscattered solar wind ENAs from the upper atmosphere (near the exobase) from Mars. By combining with results from a Monte-Carlo model by *Kallio and Barabash [2001],* we obtained the deposit of the energy to the Martian upper atmosphere of the order of $10^{10}$ eV/cm$^2$/s. It is smaller by an order than that of ultraviolet deposit.

The sub-solar hydrogen ENA jet is also the ENA flux detected by NPD. They are most probably the shocked solar wind ENA charge exchanged by the Martian exosphere near the induced magnetosphere boundary (IMB). The flux can be seen from the flank side of Mars. A rather clear decrease of the flux is seen as spacecraft moves *(Futaana et al., 2006c)*. They sometimes have a fluctuation in the flux with periods of several tens to hundreds hertz *(Grigoriev et al., 2006)*, indicating the motion or wavy structure of IMB. Associated with an interplanetary shock, a significant change in the ENA jet flux was also observed, indicating a quick shrink of IMB due to the high pressure in the solar wind *(Futaana et al., 2006d)*. These works proved that the ENA imaging will provide morphologic view and physics of IMB.

Contrary to hydrogen ENAs, we did not detect clear signatures of oxygen ENAs by NPD *(Galli et al., 2008; Futaana et al., 2011)*. Therefore, the oxygen with a form of LENA is not a significant channel of escaping atmosphere from Mars and Venus.

## ENA from the Moon

Because the Moon does not possess any atmosphere and magnetosphere, the lunar surface is exposed to the solar wind and other environment. The exposure of the surface to the environment causes changes in its physical and chemical characteristics, namely the formation of regolith. Regolith is a layer of loose, heterogeneous materials of small grain, which covers all the lunar surface. The main characteristic predicted ENA generation from regolith had been the sputtered atoms *(Futaana et al., 2006a)*. In addition, ENA production from solar wind neutralized by lunar exosphere *(Futaana et al., 2008)* or dust grain *(Collier and*

*Stubbs, 2009*) were also proposed. The scattered solar wind had not been considered as a significant ENA source, because the solar wind protons had considered to be fully absorbed (e.g. *Crider and Vondrak, 2002*), because of the high porosity of regolith.

Chandrayaan-1 spacecraft carried the first dedicated ENA sensor, CENA, to the lunar orbit (*Barabash et al., 2009*). Contrary to the above classical view, 10–20% of the solar wind protons was found scattered back as ENAs (*Wieser et al., 2009*). This is consistent with another ENA measurement by IBEX (*McComas et al., 2009*) and proton measurements from the surface of the Moon (*Saito et al., 2008*) and Phobos (*Futaana et al., 2010*).

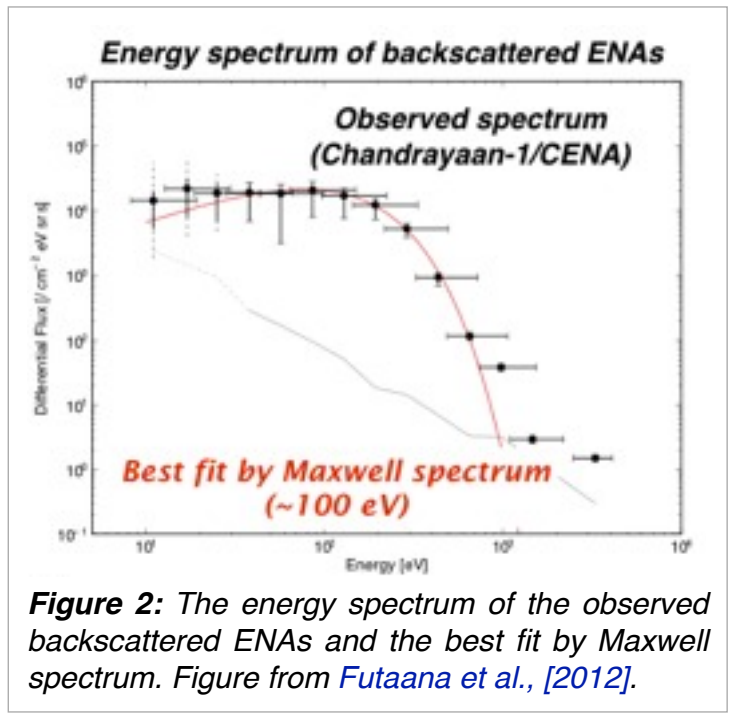

*Figure 2:* The energy spectrum of the observed backscattered ENAs and the best fit by Maxwell spectrum. Figure from *Futaana et al., [2012]*.

The contradiction between the classical assumption of full absorption and the observed 20% backscattering is yet a big open question. In addition to the high backscattering rate, particular interest is also addresses in its energy spectrum (**Figure 2**). The observed backscattered ENAs follow the Maxwell-shape spectrum with energy of ~100 eV. The spectrum suggests a thermal equilibrium of plasma, however, it is hardly plausible interpretation. Rather, the spectrum is most probably realized by a results of any randomization effects. There also found no dependency of the upstream solar wind parameters to the ENA energy spectrum, but the only exception is the linear correlation between the ENA characteristic energy in the Maxwell form (temperature) and the solar wind velocity. Why these different-dimension quantities have linear correlation is also an open question (see *Futaana et al., 2012* for details).

Whereas the thorough physical explanation is not yet available, the characteristics of the backscattered ENA energy spectrum provides unexpected but new diagnostics of the celestial surfaces, namely, the surface potential mapping (*Futaana et al., 2013*). The ENA spectrum shape (characteristic energy) provides the plasma velocity at the surface, which is modified by the surface potential. Comparison of the surface plasma velocity with the upstream plasma velocity provides the electrostatic potential. **Figure 3** shows the obtained map of the surface potential near Gerasimovic magnetic anomaly. About 150 V of the potential is formed inside the magnetic anomaly. This new technique will be applied potentially to any other celestial bodies.

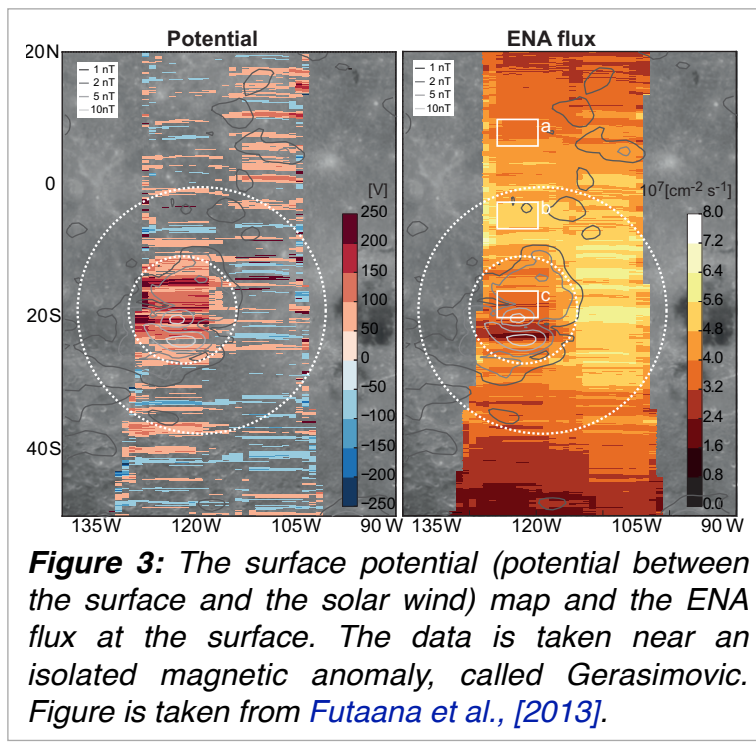

*Figure 3:* The surface potential (potential between the surface and the solar wind) map and the ENA flux at the surface. The data is taken near an isolated magnetic anomaly, called Gerasimovic. Figure is taken from *Futaana et al., [2013]*.

## Future of planetary LENA imaging

As seen above, ENA sensors are useful for plasma-planet interaction investigations. BepiColombo carry two ENA sensors to Mercury *(Saito et al., 2010; Orsini et al., 2010)*. The surface interaction with the solar wind, as of the Moon, provides the imaging of cusp region where magnetospheric ions precipitate. The potential gap measurement (*Futaana et al., 2013*) between the in situ plasma measurement (as a monitor of precipitating ions) and the backscattered ENAs (as a reference of the surface ions) could be possible. In addition, in case of the high solar wind dynamic pressure, the solar wind precipitation imaging is also plausible.